\documentclass[prb, reprint, twocolumn]{revtex4-1}
\usepackage{graphicx}
\usepackage{times}
\usepackage{upgreek}
\usepackage{color}
\begin{document}
\title{Magnetic-field-induced abrupt spin state transition in a quantum dot
containing magnetic ions}

\date{\today}

\author{M. Koperski$^{1,2}$}
\email{Maciej.Koperski@fuw.edu.pl}
\author{T. Smole\'nski$^{1}$}
\author{M. Goryca$^{1}$}
\author{P. Wojnar$^{3}$}
\author{M. Potemski$^{2}$}
\author{P. Kossacki$^{1}$}

\affiliation{
$^{1}$  Institute of Experimental Physics, Faculty of Physics, University
of Warsaw, Poland.\\
$^{2}$  Laboratoire National des Champs Magn\'{e}tiques Intenses,
CNRS-UJF-UPS-INSA, Grenoble, France\\
$^{3}$  Institute of Physics, Polish Academy of Sciences, Warsaw, Poland}

\begin{abstract}
We present the results of a comprehensive magneto-optical characterization
of single CdTe quantum dots containing a few Mn$^{2+}$ ions. We find that
some quantum dots exhibit an unexpected evolution of excitonic
photoluminescence spectrum with the magnetic field. At a certain value of
the magnetic field, specific for every quantum dot, each of the broad
spectral lines related to the recombination of various excitonic complexes
confined inside the dot transforms into a pair of narrow lines split by
several meV. We interpret this abrupt change in the character of excitonic
emission spectrum as a consequence of a transition from a non-polarized
state of the Mn$^{2+}$ spins in a low field regime to a highly (almost
fully) polarized state above the critical magnetic field. Various optical
experiments, including polarization-resolved studies, investigation of
different excitation regimes and time-resolved measurements corroborate
this scenario. However, these measurements indicate also that the observed
effect is not related or influenced by the photo-created charge carriers,
but it is rather originating from unusual spin configuration in the cluster
of Mn$^{2+}$ ions. 
\end{abstract}

\maketitle

\section{Introduction}

Semiconductor quantum dots (QDs) doped with magnetic ions constitute a
convenient platform for studying magnetism in nanoscale. Both the energy of
the system and its spin properties are encoded in the photoluminescence
(PL) signal from such QDs. Narrow spectral lines, originating from the
atomic-like states of the carriers confined in the QDs, ensure high
spectral resolution necessary for observing the detailed fine structure of
the excitonic complexes \cite{besombes2004_prl, kudelski2007_prl,
glazov2007_prb, trojnar2011_prl, trojnar2013_prb, koperski2014_prb,
kobak2014, smolenski2015_prb2, smolenski2016_natcommun, besombes2016_prb}.
The energy spectrum of such QDs is determined predominantly by the exchange
interaction between the angular momentum of the carriers and the spin of
the magnetic ions. It has been demonstrated that for QDs with a single ion
(sometimes even two ions \cite{besombes2012_prb, krebs2013_prl}) one can
distinguish isolated lines related to specific projections of the spin of
the magnetic ion (ions) on the quantization axis (given by the anisotropy
of the heavy-hole\cite{leger2005_prb}). Such direct readout of the magnetic
ion spin state has been successfully exploited not only for the studies of 
the static spin properties \cite{leger2006_prl, LeGall2011_prl}, including
the influence of an external magnetic field \cite{goryca2011_prb}, but also
dynamical processes such as relaxation \cite{goryca2015} or coherent
precession \cite{goryca2014_prl, besombes2015_prb} of a spin of a magnetic
ion incorporated into a semiconductor lattice. Additionally, more
application-oriented studies are possible, e.g., a demonstration (as a
proof of concept) of a magnetic memory based on a single spin
\cite{LeGall2009_prl, goryca2009_prl, smolenski2015_prb1}. Apart from such
extremely diluted concentration of magnetic ions, extensive efforts were
devoted to studying the properties of QDs with a large number of ions
\cite{wojnar2007_prb} (of the order of hundreds). The most interesting
findings in these QDs are related to the formation of a magnetic polaron
\cite{maksimov2000, dorozhkin2003, barman2015, nelson2015}. In particular,
the time resolved experiments revealed a significant red-shift of the
emission lines during the lifetime of excitonic complexes in QDs with
sufficiently high concentration of Mn$^{2+}$ ions (above 3~\%)
\cite{klopotowski2011_prb}.

With this work we aim to fill the gap in the research of magnetic QDs by
exploring the regime of intermediate concentration of magnetic ions, when
on average only a few ions are present in a single QD. For some of such
QDs, in which the Mn$^{2+}$ ions appear to be located at the QD's edge, we
found a new type of the evolution of the PL spectra with the magnetic field
in Faraday configuration, indicating a rapid transition of the spin state
of the magnetic ions at a certain, critical value of the magnetic field
ranging from 3 to 10~T for various QDs. Such a behavior of the system of
spins in a QD comes as a surprise, since it strongly deviates from the
typical magneto-PL data for the magnetic QDs studied so far. A firm
explanation of the observed transition from a non-magnetic to a highly
(almost fully) polarized spin state of the magnetic ions remains to be
established. However, the comprehensive study of this effect by means of
various spectroscopic tools presented in details here provides a solid
ground for possible interpretations.

\section{Samples, experimental setup and basic optical characterization}

The samples used in our experiments were grown by the molecular beam
epitaxy (MBE). They contained a single layer of CdTe QDs embedded in ZnTe
barrier. The doping with Mn atoms was introduced into the QD formation
layer and adjusted so that a significant number of QDs would contain a few
Mn$^{2+}$ ions. The initial estimation of the Mn$^{2+}$ ions concentration
was based on the calibration of the temperature of the Knudsen effusion
cell (related to a flux of the Mn atoms in the MBE growth process). In the
case of the most intensively investigated sample in this work, the Mn atoms
were introduced to one out of six deposited CdTe monolayers, and the
estimated Mn concentration in this monolayer yields 0.4\%. Therefore the
effective concentration of the Mn$^{2+}$ ions in the QDs can be estimated
to be $~$0.07\%. This corresponds to a presence of 2~--~7~Mn$^{2+}$ ions in
a single QD (depending on the assumed size of a QD within reasonable
constrains established, e.g., from the atomic force microscope studies
\cite{wojnar2008}).

The samples were first investigated with magneto-PL experiments, which were
followed by comprehensive time-resolved studies. The high magnetic field
experiments (up to 29 T) were performed in a resistive magnet. In this
setup the samples were placed inside a probe filled with helium exchange
gas (cooled down to about 10 K) and equipped with piezo-stages allowing
x-y-z sample positioning with sub-micrometer precision. A fiber-based
optical system was used for the excitation of the sample (typically with
argon laser lines 488 or 514~nm) and for the detection of the spectrally
resolved PL signal with a charge-coupled device (CCD) camera. The laser
spot diameter in this setup was $\sim$1~$\upmu$m. 

For time-resolved experiments a split-coil superconducting magnet with a
direct optical access was used. The sample was cooled down with pumped
liquid helium to 1.7~K. An immersive reflective microscope allowed a
focalization of the laser beam down to a sub-micrometer spot. In this
setup, apart from further PL characterization of the samples, the temporal
evolution of the PL spectra (with overall resolution of $\sim$10~ps) was
measured with a streak camera. Also, the effects of an introduction of a
dark period in the excitation (tens of microseconds duration) were studied
by using an electrically controlled laser with fast (single nanoseconds of
rise/fall times) turn on/off module for excitation and a CCD camera
equipped with a gated microchannel plate for detection.

\begin{figure}
\includegraphics[width=85mm]{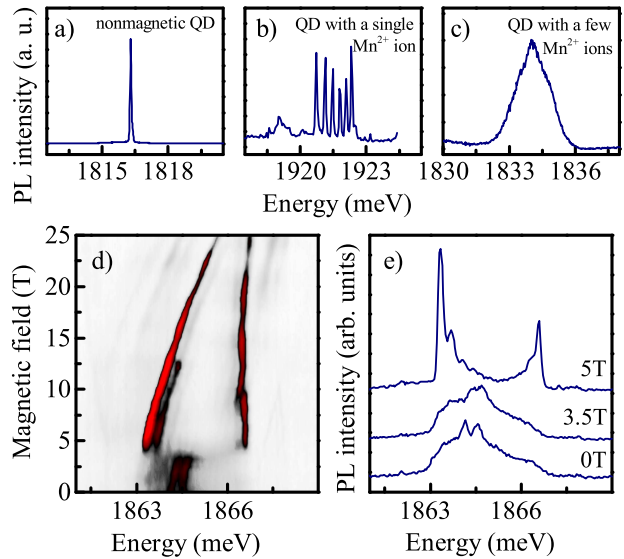}
\caption{ (a~--~c) Example PL spectra showing a neutral exciton transition
without the magnetic field for three selected QDs from the same sample. The
spectra correspond to (a) a nonmagnetic QD, (b) a QD with a single
Mn$^{2+}$ ion and (c) a QD with a few Mn$^{2+}$ ions. For some QDs with a
few Mn$^{2+}$ ions the evolution of the spectrum with the external magnetic
field applied in Faraday configuration reveals an abrupt transformation of
the PL lines, as seen in a color-scale map of magneto-PL spectra presented
in (d). From a single broad feature the PL line transforms into two
separate narrow lines, which is clearly illustrated by the PL spectra
measured at selected values of the magnetic field (e). \label{fig:spec}}
\end{figure}

The studies presented here concern the optical properties of single QDs.
Although the large concentration of QDs in our samples (of the order of
$10^{10 }$~cm$^{-2}$ [\onlinecite{wojnar2008}]) results in about
$\sim$100~QDs under the laser spot, one can easily distinguish sets of
separate lines from single QDs in the low-energy tail of the broad PL band.
As the concentration of the Mn$^{2+}$ ions in our samples is rather low, it
is still possible to recognize the PL signal from single nonmagnetic QDs
with a typical pattern of individual narrow lines (linewidth below
100~$\upmu$eV) corresponding to the recombination of different excitonic
complexes. At the same time, the QDs with a single Mn$^{2+}$ ions may be
identified, based on the characteristic splitting of the excitonic lines
(sixfold in the case of the neutral exciton). However, the PL spectra of
the majority of single QDs in the sample are composed of significantly
broadened separate lines. Example spectra for these three cases are
presented in Figs. \ref{fig:spec}a~--~c. In the view of the
characterization of our samples discussed so far and the magneto-PL data
presented later on, we interpret the observed broadening of the lines as
resulting from the presence of a few Mn$^{2+}$ ions in most of the QDs. The
origin of the broadening is related to the thermal spin fluctuations in the
cluster of Mn$^{2+}$ ions in the QD, which significantly extends the energy
space available for a particular excitonic complex. As the large number of
possible spin configurations grows exponentially with the number of
Mn$^{2+}$ ions, the detailed fine structure may no longer be resolved in
the case of three or more Mn$^{2+}$ ions embedded in the QD. Therefore, in
the PL spectra the excitonic transitions are seen as broader features.

\section{Magneto-luminescence studies}
The magneto-luminescence studies of QDs exhibiting a broad emission lines
reveal an unexpected behavior of the spin of the Mn$^{2+}$ ions under the
influence of an external magnetic field. For a certain number of
investigated QDs (roughly about 20~$-$~30\% for different samples) an
abrupt transformation occurs in the character of the PL spectrum: a single
broad PL line splits into two narrow components at a certain critical value
of the magnetic field, as presented in Figs.~\ref{fig:spec}d~--~e. Above
the critical field, the two narrow lines follow the magnetic field
evolution similar to the one known from the studies of excitonic
transitions in nonmagnetic QDs. In particular, the energies of both narrow
lines evolve exclusively due to the linear excitonic Zeeman effect and
quadratic diamagnetic shift. Moreover, the linewidth of the  two emission
above the critical field is similar to the one observed for nonmagnetic QDs
and remains roughly constant with the further increase of the magnetic
field, as  one could expect for (Cd,Mn)Te QDs in the regime of high (almost
full) spin polarization of the Mn$^{2+}$ ions. The two lines tend to cross
at high magnetic field, which remains above the experimental limitation of
the present study. That crossing would correspond to the compensation of
the excitonic Zeeman effect and the giant Zeeman effect of the Mn$^{2+}$
ions, which shift the energies of excitonic spin states in the opposite
directions for (Cd,Mn)Te material. Based on these observations, we
interpret the transformation of the PL lines as a transition from a
non-polarized to a highly polarized spin state of the Mn$^{2+}$ ions.

\begin{figure}
\includegraphics[width=85mm]{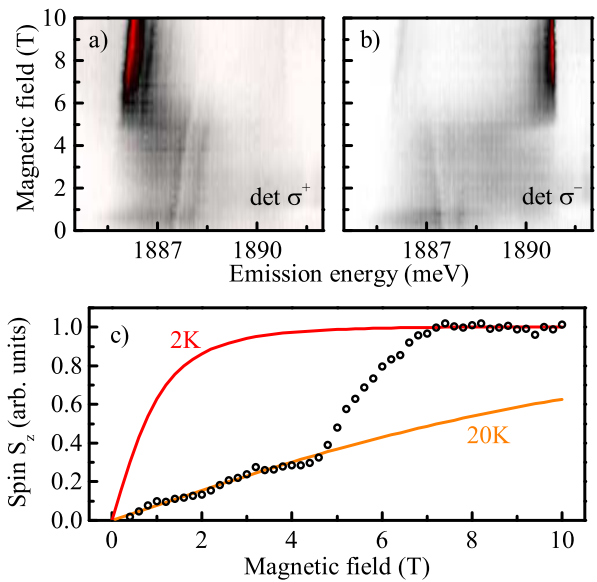}
\caption{(a~--~b) The polarization-resolved magneto-PL maps of a neutral
exciton for a QD with a few Mn$^{2+}$ ions (sample temperature of 1.7~K)
exhibiting an abrupt transformation. The maps were measured in (a)
$\sigma^+$ and (b) $\sigma^-$ polarization of detection. (c) The evolution
of the mean spin of the Mn$^{2+}$ ions obtained from these data (see main
text). The solid lines represent the Brillouin functions for 5/2 spin at
effective temperatures of 2~K and 20~K.\label{fig:pol}}
\end{figure}

In order to quantitatively characterize the transformation, we turn to the
analysis of the polarization-resolved magneto-PL spectra of the neutral
exciton, which are presented in Figs.~\ref{fig:pol}a~--~b. 
We tentatively assume that the PL signal measured in a given circular
polarization reflects directly the probability distribution of total spin
of the Mn$^{2+}$ ion system. In such a case, the calculation of the first
moment of the PL intensity distribution at a certain value of the magnetic
field yields a quantity reflecting, in the first approximation, the
magnetization of the Mn$^{2+}$ ions in a QD. The circular polarization
resolution in detection allows us to account for the excitonic relaxation
processes, which for some QDs are clearly manifested in zero-field spectra
as non-symmetric shape of the excitonic lines. Therefore, we use a
difference between the magnetization of Mn$^{2+}$ ions calculated from the
PL spectra detected in opposite circular polarizations to obtain a value
unperturbed by the excitonic relaxation. Such value is presented in
Fig.~2c as a function of the magnetic field. A clear transition is seen,
starting at $\sim$5~T and progressing until $\sim$7~T, when a saturation
value is reached. For the reference, the evolution of the magnetization is
compared with the two Brillouin function curves for a 5/2 spin at effective
temperature of 2~K (close to the helium bath temperature equal to
$\sim$1.7~K) and 20~K. The values of the temperature are chosen so that the
Brillouin curves match the magnetization data in the higher and lower field
regimes (after and before transition, respectively). Even though the
Brillouin function, originating from a mean field approximation applied to
a paramagnetic system, may not be expected to accurately describe the spin
state of a few interacting Mn$^{2+}$ ions, the comparison suggests an
existence of a mechanism suppressing the spin polarization of the Mn$^{2+}$
ions in the low field regime, which is abruptly switched off at the field
when the transition is seen.

\begin{figure}
\includegraphics[width=89mm]{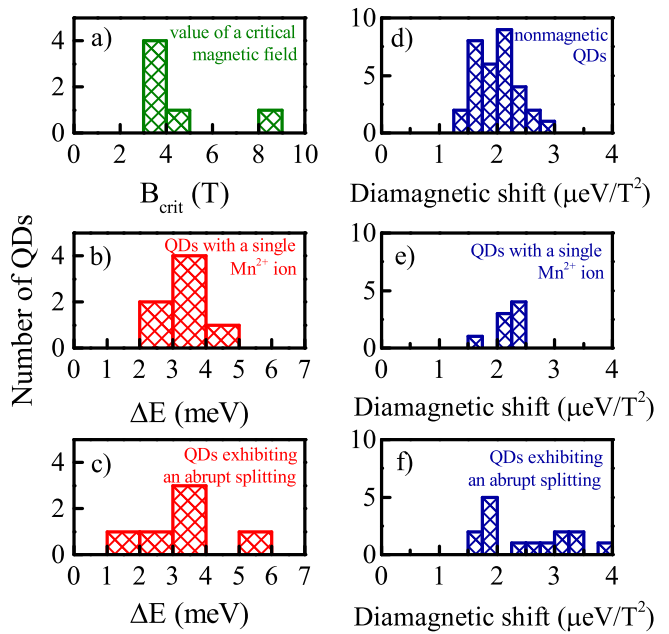}
\caption{(a) Histogram presenting the value of a critical magnetic field,
which induces an abrupt transition of the PL lines from QDs with a few
Mn$^{2+}$ ions. (b, c) Zero-field effective exchange splittings of the
neutral exciton measured for the QDs containing single Mn$^{2+}$ ions (b)
and for the dots with a few Mn$^{2+}$ ions (c). (d -- f) Histograms of
diamagnetic shift coefficients determined for (d) nonmagnetic QDs, (e) QDs
with single Mn$^{2+}$ ions and (f) for the QDs with a few Mn$^{2+}$ ions
exhibiting a magnetic-field-induced transition. The data clearly show that
in the latter case the diamagnetic shift is higher, which strongly
indicates that the QDs exhibiting the transformation are of larger size.
\label{fig:stat}}
\end{figure}

An analysis of the magnetic field evolution of the PL spectra for several
QDs exhibiting a transition allows us to establish basic properties of the
cluster of Mn$^{2+}$ ions in such specific QDs. First, an important
parameter describing the phenomenon is the value of a critical magnetic
field at which the transition is seen. Its values for several QDs are
presented in Fig.~\ref{fig:stat}a. The critical value of the magnetic field
is a particular property of a specific QD and typically ranges from 3~T up
to 5~T, however for some QDs can reach as high as 10~T. Second, we can
determine an effective zero-field exchange splitting between
$\upsigma^+$/$\upsigma^-$ exciton interacting with highly spin-polarized
Mn$^{2+}$ ions by extrapolating the evolution of the two narrow lines above
the critical field down to the zero magnetic field. Such a splitting
depends on the number of Mn$^{2+}$ ions in the QD, their position in the
lattice structure and the shape of the excitonic wave function governed by
the geometry of the QD, all of which finally determine the wave function
overlap between the Mn$^{2+}$ ions and the exciton. It is interesting to
compare the average value of this splitting with a typical value of the
zero-field exchange splitting for the QDs doped with single Mn$^{2+}$ ions
from the same sample, which is directly available as a splitting between
the two outermost lines in a neutral exciton sextuplet. Relevant
statistical data are presented in Fig.~\ref{fig:stat}b~--~c in form of
histograms for QDs with a single and with a few Mn$^{2+}$ ions,
respectively. On average the value of the splitting in both cases is
roughly the same, which suggests that for the QDs exhibiting the transition
the Mn$^{2+}$ ions are located rather at the perimeter of the QD than in
its center (one has to note, however, than in case of the statistics for
the QDs with a single Mn$^{2+}$ ion, the average value is shifted towards
higher energy by our choice of QDs with clearly resolvable six lines). An
additional factor that further reduces the Mn-exciton wave function overlap
is related to a larger size of the QDs with a few Mn$^{2+}$ exhibiting a
transition, which is revealed by the higher value of the diamagnetic shift
for these QDs as compared to the QDs without or with a single Mn$^{2+}$ ion
(Fig.~\ref{fig:stat}d~--~f).

\begin{figure}
\includegraphics[width=89mm]{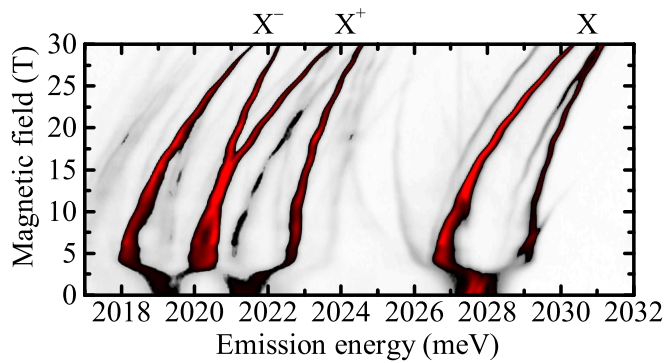}
\caption{The magnetic field evolution of a PL spectrum of a single
CdTe/ZnTe QD with a few Mn$^{2+}$ ions presented in a wider spectral range.
The data show that the PL line transformation appears for all excitonic
complexes visible in the QD PL spectrum. In particular, both neutral and
charged complexes reveal the same kind of transformation at exactly the
same value of the critical magnetic field.   \label{fig:magnetoPL}}
\end{figure}

At this point a question arises about the origin of the observed effect:
whether the transition is caused by some process mediated by photo-created
carriers, which could be controlled by means of optical techniques or is it
an intrinsic property of the cluster of the magnetic ions due to a
particular spin configuration defined by their position in the crystal
lattice, mutual exchange interaction and interaction with the residual
carriers. In order to address these two possibilities, we first analyze the
magneto-PL spectra in a wider energy range, covering the emission of
different excitonic complexes. Fig.~\ref{fig:magnetoPL} illustrates the
magnetic field evolution of a spectrum for a selected QD exhibiting a
transformation. The complete spectrum of a CdTe QD consists of a set of
lines, which form a characteristic pattern enabling the identification of
particular excitonic complexes. Hence, it is possible to recognize the
emission of both neutral and charged states. The lines related to the
neutral exciton and negatively/positively charged excitons are indicated in
the figure. Importantly, the same kind of transformation is seen for all
excitonic lines. Particularly interesting in this context are the trion
states, which are in fact composed of 3 carriers, however due to the
singlet state of a pair of carriers of the same type the trions effectively
interact with the Mn$^{2+}$ spins in the same way as a single minority
carrier (the electron in case of positive trion and the hole in the case of
negative trion). Apparent equal robustness of the observed transformation
regardless of the presence of an electron, a hole as well as an
electron-hole pair suggests that there is no link between observed
excitonic complex and the actual origin of the effect.

\begin{figure}
\includegraphics[width=89mm]{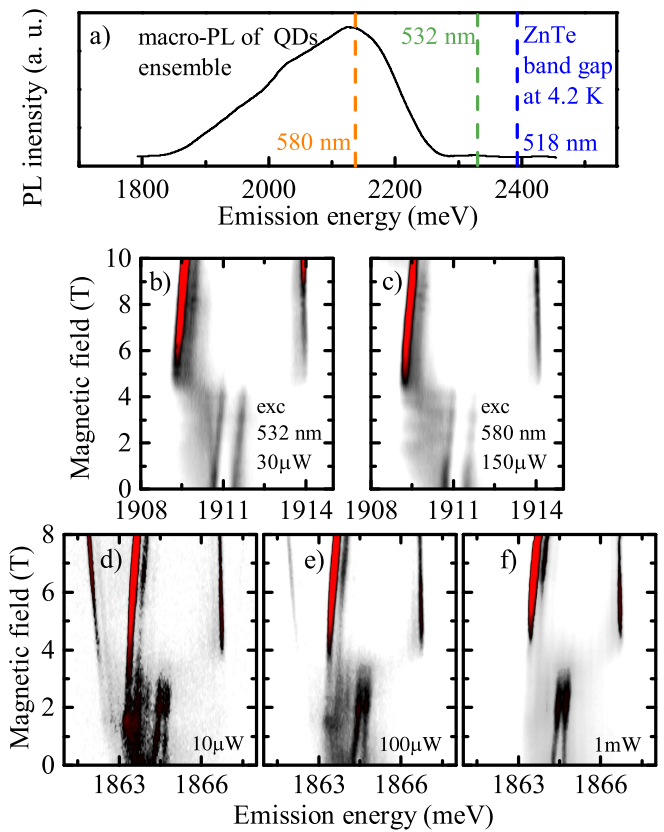}
\caption{The presence of magnetic-field-induced abrupt transformation of
the PL line for CdTe/ZnTe QDs with a few Mn$^{2+}$ ions is found to be
independent of the laser excitation conditions. This includes two different
regimes of the excitation energy schematically depicted in panel (a): just
below the barrier band gap with 532 nm diode laser and quasi-resonant with
580 nm tunable dye laser.  The color-scale maps presenting the magneto-PL
spectra obtained under both excitation regimes (b -- c) confirm that the
character of the abrupt transformation remains unaffected by the choice of
the laser energy. Additionally, no particular influence of the laser
excitation power on the transformation is observed, as seen in the data
presented for (d) 10~$\upmu$W, (e) 100~$\upmu$W and (f) 1~mW laser
excitation power of 514~nm Ar laser line. \label{fig:exc}}
\end{figure}

The type of transformation of the PL lines in the magnetic field that we
describe here is not known to appear in other magnetic systems. Seemingly
similar deviations from the typical giant Zeeman effect were observed for
CdTe/ZnTe quantum wells \cite{scherbakov1999} or QDs \cite{clement2010}
with high concentration of magnetic ions. These deviations were interpreted
as related to the influence of carriers trapped in a wetting layer, which
can effectively interact with the Mn$^{2+}$ ions only in the low magnetic
field regime. More specifically, at low fields the spin-flip processes
between the angular momentum of such carriers and the spin of the Mn$^{2+}$
ions enable the transfer of energy and may lead to an increased temperature
of the Mn$^{2+}$ spins. However, a similar scenario is highly unlikely in
our case for the reasons discussed below.

First, we verify such hypothesis experimentally by changing the excitation
conditions (energy and power). A typical macro-PL spectrum of an ensemble
of QDs from our samples is presented in Fig.~\ref{fig:exc}a. It covers a
spectral range roughly from 1850 meV to 2250 meV. A potential wetting layer
in our sample (however not appearing in the PL signal) would reside at
higher energies but below the ZnTe (barrier) bandgap. In
Fig.\ref{fig:exc}b~--~c we present the magnetic field evolution of a PL
line for a QD exhibiting a rapid transformation measured under two
excitation regimes: just below the band gap (with a 532 nm diode laser) and
quasi-resonant (with a tunable rhodamine dye laser set for 580 nm). While
the creation of carriers in a potential wetting layer is feasible for the
532 nm excitation, the quasi-resonant excitation excludes such possibility.
The presence of equally robust PL line transformation for both regimes of
excitation directly confirms that the carriers in the potential wetting
layer cannot be responsible for the observed effect. Moreover, any heating
mechanism should be strongly dependent on the excitation power. For
example, in the case of QDs with single Mn$^{2+}$ ions, the heating of the
ion spin by the laser excitation is known to be very efficient with the
corresponding effective temperatures of the Mn$^{2+}$ ion reaching tens of
K for sufficiently high excitation powers \cite{besombes2004_prl,
kobak2014, smolenski2015_prb2, papierska2009}. Here, we do not observe any
significant change in the nature of the transformation by varying the laser
excitation power by two orders of magnitude, as shown in
Fig.~\ref{fig:exc}d~--~f.

Second, the magnetic field needed to suppress the potential heating
mediated by the carriers in the wetting layer is proportional to the
concentration of the magnetic ions. Such concentration in our samples is
almost an order of magnitude lower than in the samples used in the previous
reports. This is revealed, for instance, by the energy splitting between
the two $\sigma^+/ \sigma^-$ polarized lines in the magnetic field that
highly polarizes the magnetic ions spins. The splitting equals 2~--~4 meV
in our samples as compared to the splitting of about 20~meV observed for
the samples used before. Consequently, in our case the heating could be
effective only for the magnetic fields remaining in the range of tens of
mT, which are much smaller that the values of critical magnetic fields
3~T~--~10~T at which the rapid transformation of the PL lines appears. This
observation finally shows that the transformation cannot be due to the
invoked heating mechanism.

\section{Evidence for a steady spin state of Mn$^{2+}$ ions in
time-resolved experiments}

\begin{figure}
\includegraphics[width=89mm]{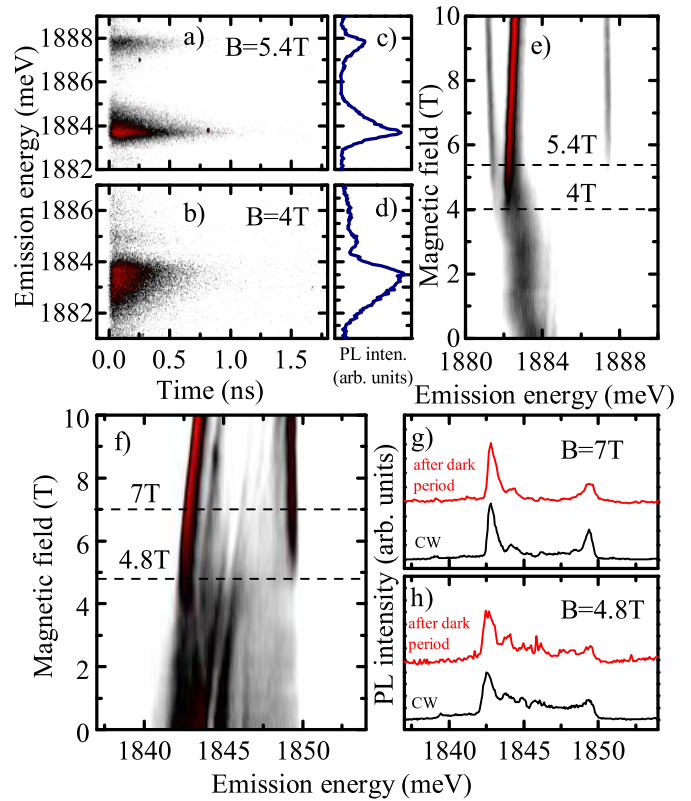}
\caption{The temporal traces of the neutral exciton PL line measured for
the QD containing a few Mn$^{2+}$ ions with a streak camera in the two
values of the magnetic field: (a) 4T and (b) 5.4T. The respective
time-integrated spectra are shown in (c) and (d) for comparison. The
critical value of the magnetic field for this QD is equal to about 5 T, as
seen in the magnetic field evolution of the PL spectrum (e). For the second
time-resolved experiment the neutral exciton PL spectrum after a dark
period was compared with the one obtained under CW excitation. Another QD
with a transition at 5 T was selected, as seen in (f). The comparison of
the two spectra measured in the magnetic field (g) below and (h) above the
transformation indicate no influence of the introduced dark period on the
character of the transformation.  \label{fig:TR}}
\end{figure}

Valuable information relevant for establishing the origin of the abrupt
field-induced transformation of the PL lines in the studied QDs may come
from the time-resolved photoluminescence measurements. Such experiments can
provide deeper insight into the character of the transformation and help to
distinguish between the intrinsic static spin properties of the Mn$^{2+}$
ions and dynamical processes related to the capture/recombination of the
photo-created carriers in the QD.  In fact, the data obtained from the
time-integrated PL experiments do not exclude the possibility that the spin
of the Mn$^{2+}$ ions becomes highly polarized at the critical value of the
magnetic field due to the interaction with excitonic complexes. One
approach to verify such scenario is to investigate the PL transients. Such
measurements would reveal the possible variations of the Mn$^{2+}$ ions
magnetization in the timescale  shorter or comparable to the excitonic
lifetime ($\sim$300~ps for CdTe/ZnTe QDs). In order to trace the temporal
evolution of the PL spectra with high resolution ($\sim$10 ps) we use a
streak camera. As the line transformation in the magnetic field is observed
for all excitonic complexes, we focus on a neutral exciton line. The
temporal evolutions of the neutral exciton PL line measured for a selected
QD in the two values of the magnetic field (4~T and 5.4~T) are presented in
Fig.~\ref{fig:TR}a~--~b. The corresponding time-integrated PL spectra
obtained in the same fields are shown in Fig.~\ref{fig:TR}c~--~d. The
values of the magnetic field were chosen to be just below and just above
the transformation (the evolution of the PL spectrum with the magnetic
field is presented in Fig.~\ref{fig:TR}e). The temporal profiles in both
regimes of the magnetic field do not indicate any transformation of the
spectral lines during the lifetime of a single excitonic complex. 

The measurement of the PL decay does not entirely exclude that the observed
transformation of the PL lines is related to the photo-created carriers.
Another possibility is that the transformation originates from a stationary
state established by a series of multiple capture-recombination events. The
second time-resolved experiment was designed to verify such hypothesis. In
these measurements we used a continuous-wave (CW) 405 nm laser with a
module of fast, electrically controlled turn on/turn off system, which
enabled us to introduce a dark period (50 $\upmu$s) in the excitation. The
dark period of such duration is significantly longer than the
characteristic spin relaxation times of clusters of Mn$^{2+}$ ions embedded
in CdTe crystal \cite{goryca2015}. Consequently, at the moment of
excitation relaunch the system of Mn$^{2+}$ ions may be considered as fully
thermalized. During the excitation period (1 $\upmu$s) the PL spectrum was
recorded with a gated CCD camera in a series of temporal windows of 30 ns,
allowing us to probe the stationary state of the Mn$^{2+}$ ions in darkness
(i.e., absence of excitation). Again, a QD exhibiting a rapid
transformation was selected (the magneto-PL map shown in
Fig.~\ref{fig:TR}f) and the experiment was performed at two values of the
magnetic field (4.8~T and 7~T), chosen in the middle and just above the
transformation. In Fig.~\ref{fig:TR}g~--~h the two PL spectra are compared:
the one under CW excitation and the second one recorded during 30~ns time
window just after the dark period. No particular difference observed
between these two spectra allows us finally conclude that the field-induced
transformation of the excitonic lines is not mediated by the photo-created
carriers. This in turn strongly suggests that the origin of the effect is
related to the intrinsic properties of the magnetic cluster of Mn$^{2+}$
ions.

\section{Summary}

Based on the magneto-optical studies of CdTe/ZnTe QDs doped with a few
Mn$^{2+}$ ions we have uncovered an abrupt transition from a non-polarized
to a highly polarized state of the magnetic ion system inside the dot,
which occurs upon application of the magnetic field in the Faraday
geometry. The observed effect of the transformation of the PL lines has
proven to be insensitive to the conditions of excitation, both in
time-integrated and time-resolved domain. Moreover, the transition turned
out to remain unperturbed, when probing the spin state of the Mn$^{2+}$
ions with different excitonic complexes, both neutral and charged.
Therefore, all our findings point towards the conclusion that the origin of
the effect is not related to the influence of photo-created charge
carriers. At present we consider mainly two scenarios. One possibility is
that the Mn$^{2+}$ ions were incorporated into the crystal lattice so close
to each other that their mutual exchange interaction is significant. In
such a case they could form a highly frustrated system exhibiting a
transition in the magnetic field, in a similar manner as, i.e., magnetic
ions in single molecules, which can be found in particular configurations
such as triangular clusters \cite{luzon2008}. On the other hand, even
distant Mn$^{2+}$ ions can interact with each other via exchange
interaction with a carrier\cite{krebs2013_prl}, which can stochastically
occupy the QD in the absence of optical excitation. Such an interaction can
have non-trivial character, especially in the case of the hole, with its
intrinsic anisotropy additionally complicated by possible effects of spin
textures\cite{abolfath2012, stirner1995}.

In view of the presented results, the PL measurements cannot provide
definitive answer about the origin of the effect. Therefore, in the future
studies one should include other techniques, such as optically detected
magnetic resonance or non-optical magnetization measurements.

\section*{Acknowledgements}
We thank T. Kazimierczuk for fruitful discussions. This work was supported
by National Science Center (NCN) Projects No. DEC-2012/07/N/ST3/03665 and
No.~DEC-2011/02/A/ST3/00131.

\end{document}